\documentclass{llncs}
\usepackage{graphicx}
\usepackage{amsmath}
\usepackage{algorithm}
\usepackage{algorithmic}

\begin{document}
\frontmatter          
\pagestyle{headings}  
\mainmatter   

\title{Energy-Efficient Offloading in Mobile Edge Computing with Edge-Cloud Collaboration\thanks{This work has been accepted by the 18th International Conference on Algorithms and Architectures for Parallel Processing (ICA3PP 2018). The copyrights are held by authors and the corresponding copyright holders. This document is only for quick dissemination of research findings. This document is not the final vision of the conference paper. For detail, please see the original conference paper when it is online. Please cite: Xin Long, Jigang Wu, Long Chen, Energy-Efficient Offloading in Mobile Edge Computing with Edge-Cloud Collaboration, 18th International Conference on Algorithms and Architectures for Parallel Processing November 15-17, 2018, China.
}}
%
%

\author{Xin Long \and
Jigang Wu\and
Long Chen}
%
%
\institute{School of Compt. Sci. \& Tech.\\
Guangdong University of Technology \\
Guangzhou, China\\
\email{longyiyuan@outlook.com, asjgwucn@outlook.com, lonchen@mail.ustc.edu.cn}}

\maketitle

%
%
\begin{abstract}
Multiple access mobile edge computing is an emerging technique to bring computation resources close to end mobile users. By deploying edge servers at WiFi access points or cellular base stations, the computation capabilities of mobile users can be extended. Existing works mostly assume the remote cloud server can be viewed as a special edge server or the edge servers are willing to cooperate, which is not practical. In this work, we propose an edge-cloud cooperative architecture where edge servers can rent for the remote cloud servers to expedite the computation of tasks from mobile users. With this architecture, the computation offloading problem is modeled as a mixed integer programming with delay constraints, which is NP-hard. The objective is to minimize the total energy consumption of mobile devices. We propose a greedy algorithm as well as a simulated annealing algorithm to effectively solve the problem. Extensive simulation results demonstrate that, the proposed greedy algorithm and simulated annealing algorithm can achieve the near optimal performance. On average, the proposed greedy algorithm can achieve the same application completing time budget performance of the Brute Force optional algorithm with only 31\% extra energy cost. The simulated annealing algorithm can achieve similar performance with the greedy algorithm.

\end{abstract}
\section{Introduction}
The recent tremendous growth of various wireless devices and diverse applications has brought the challenge in wireless systems. Since the proliferation of smart mobile devices and wearable sensors, mobile traffic and computation tasks have increased dramatically. Therefore, cloud computing \cite{barbera2013offload} as well as 5G communication \cite{chen2018brains,dhillon2012modeling} has been proposed to deal with this challenge in the big data era. Despite the potential in data storage and analysis, cloud computing cannot fulfill the growing application requirements such as low latency and context awareness. Multiple-access mobile Edge Computing (MEC) \cite{hu2015mobile} that serves as a complement for cloud computing can potentially overcome the weakness of mobile cloud computing by offloading computation intensive tasks at the edge of wireless networks \cite{chen2018tarco}. 

Task allocation and computation resource assignment are crucial to MEC, especially in the presence of an application with a large number of delay sensensitive subtasks. For example, on-line gaming for recreation or face recognition for security purposes. Those tasks should be handled in time taking the finite bandwidth and limited computation resources into consideration. The offloading problem that taking into consieration the above factors jointly are usually mixed integer programming problems which are non-convex and NP-hard \cite{Chen2016Joint} \cite{chen2017joint}. Among the task allocation and resource assignment schemes, energy optimization is one of the key factors that affect the performance of the computaton resource limited mobile devices. That's because the energy consumption of mobile devices would exponentially grow when there are multiple complex tasks on the devices.

Earlier works on energy optimization for MEC, such as \cite{Bi2018Computation,lyu2018energy}, assumed unlimited energy supply of edge servers. Bi et al. \cite{Bi2018Computation} addressed the computation rate maximization problem in wireless powered MEC networks. Mobile devices can harvest energy from the cellular base station that with an MEC server. The original problem was non-convex and a decoupled optimization with coordinate descent method was proposed to solve the proposed problem. Lyu et al. in \cite{lyu2018energy} studied the total energy consumption of multiple devices with latency constraints. The problem was modeled as a mixed-integer programming, followed by a dynamic programming algorithm based on Bellman equation. More recent researches \cite{chen2017energy,Kao2015Hermes} have been focused on delay minization with energy or budget constraints of edge servers. Chen et al. \cite{chen2017energy} carried out with a novel multi-cell MEC architecture where edge devices such as base stations can cooperate with remote server on task execution. Considering the ON/OFF nature of edge servers, they used Lyapunov optimization
technique to obtain optimal decisions on task offloading. Considering task dependency, Kao et al. \cite{Kao2015Hermes} presented Hermes, aiming at minimizing total execution time of tasks with user budget constraints. 

\begin{table}[htb]
	\caption{Comparison between existing works and this work.}\label{tb:comp}
	\begin{tabular}{|l|l|l|l|l|l|}
		\hline
		Existing works &\cite{Bi2018Computation} & \cite{chen2017energy}&Hermes \cite{Kao2015Hermes}&\cite{lyu2018energy}& This work\\
		\hline
		Task Dependency &  No & No& Yes& No& Yes\\
		Edge-Cloud Collaboration &  No & No & No &No & Yes\\
		Energy Constraint of Users & No & Yes& Yes&No & Yes\\
		Server Utility Constraint &No & No &No&No & Yes\\
		Objective & Computation Rate &  Delay& Delay&Energy & Energy\\
		\hline
	\end{tabular}
\end{table}

Based on the literature reviews, task dependency was not properly investigated by \cite{Bi2018Computation,chen2017energy,lyu2018energy}, which is important for real deployment. Although task dependency was used in the model by \cite{Kao2015Hermes}, authors in \cite{Kao2015Hermes} merely neglected the influence of remote cloud servers. Moreover, all the above works assume the remote cloud server can be viewed as a special edge server or the edge servers are willing to cooperate. In real senarios, the remote cloud server has higher computation capability than the edge server and the transmission delay between edge cloud and remote server cannot be neglected when designing proper offloading schemes. Take face recognition as an example. The feature extraction tasks for face images obtained by individual mobile devices can be offloaded to edge servers while the machine learning and face recognization, i.e., image matching tasks can be executed on the remote cloud servers. Therefore, with edge-cloud cooperation, the target faces can be detected with certain bounded delay for distributed mobile devices.

In this work, we investigate computation offloading decision and resource allocation problem with given delay requirements of mobile applications. The objective is to minimize sum energy consumption of mobile devices. Different from above works, we take edge-cloud cooperation into account, which being new challenges for the energy optimization problem. Since there are heterogeneous network resources, it is necessary to determine which the computation tasks should be done at remote clouds, proccessed at edge servers or local mobile devices.  From the perspective of edge and remote cloud servers, their service for mobile devices should be compensated for the cost of execution and their profits should be guaranteed. Since the tasks of one application is delay bounded, how to handle edge-cloud cooperation with user budget constraints should be carefully designed.

The main contributions of this paper can be summarized as follows:
\begin{itemize}
	\item  A novel edge-cloud cooperation architecture is proposed in wireless heterogeneous network with edge servers deployed at small-cell base stations and remote cloud servers connected to the macro-cell base station. The edge server can hire remote edge servers to process some of the tasks originated from mobile devices.
	\item  The offloading problem is modeled as a mixed integer non-linear programming, which is NP-hard. We then propose a greedy algorithm as well as a simulated annealing algorithm to effectively solve the problem.  
	\item  To provide incentive for edge servers, we propose a pricing scheme with virtual currency from mobile users to edge servers and remote cloud servers for the dedication of servers serving mobile users.
	\item  Extensive simulation results demonstrate that, the proposed greedy algorithm and simulated annealing algorithm can achieve the near optimal performance. On average, the proposed greedy algorithm can achieve the same application completing time budget performance of the Brute Force optional algorithm with only 31\% extra energy cost. The simulated annealing algorithm can achieve similar performance with the greedy algorithm.
\end{itemize}

The remainder paper is organized as follows. System model and computation model are presented in Section 2. Section 3 presents the problem formulation. The proposed algorithms is described in Section 4. Section 5 presents the performance evaluation. Section 6 concludes this paper with future remarks.

\section{System Model  and Computation Model }
This section firstly describes the system model and formulates the offloading problem for energy saving with local computing, edge computing and the collaboration between edge and cloud servers.
\subsection{System Model}
\begin{figure}
	\centering
	\includegraphics[width=4.5in]{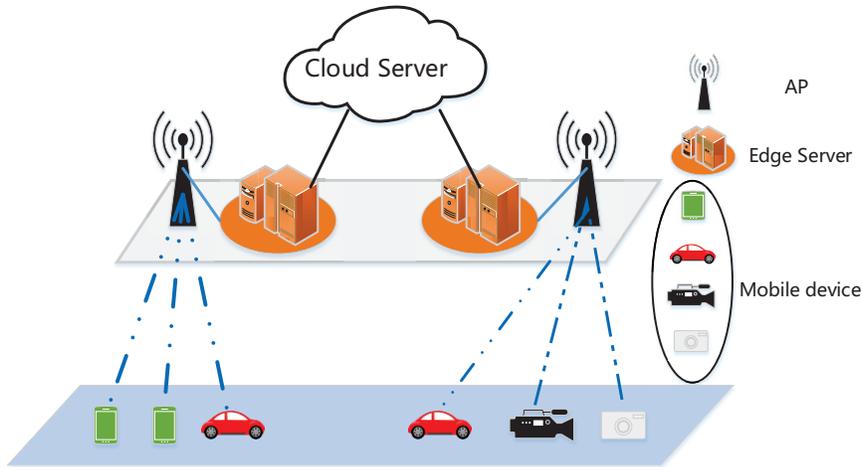}
	\caption{System Architecture} \label{systemodel}
\end{figure}
As shown in Fig. \ref{systemodel}, each edge server is located at the access point (AP) \cite{chen2018quick} which is also being attached by multiple mobile devices. The edge server is deployed at the AP and is linked to the remote cloud via high speed fiber links. Let $U$ be the set of mobile devices, We assume that there are $M$ mobile devices. Therefore, we have $U = \{u_{1}, u_{2}, u_{3}, \cdots, u_{M}\}$, where $M \geq 1$. Meanwhile, there is a set $T_m$ subtasks on the $m$-th mobile device, which cloud be denoted as $T_m= \{\tau_{m,1}, \tau_{m,2}, \tau_{m,3}, \cdots, \tau_{m,N}\}$, where $ N \geq 0 $.

Next, we will introduce the communication and computation models for mobile devices, edge servers and remote cloud in detail. 

\begin{table}[!h]
	\caption{Basic Notations}
	\begin{center}
		\begin{tabular}{c| p{9cm}}	
			\hline
			{\bf Notation } & {\bf Descriptions}\\
			\hline
            $M$ & Number of mobile devices \\
            $N$ & Number of subtasks \\
            $D_{m,n}$ & Data size of subtask $n$ on mobile device $m$ \\
            $W_{m,n}$ & Workload of subtask $n$ on mobile device $m$ \\
            $R_{m,n}$ & Uplink data rate for subtask $n$ of mobile device $m$\\
            $t_{m,n}^{t}$ & Time spent when sending subtask $n$ of device $m$ to edge server \\
            $t_{m,n}^{r}$ &Time spent when sending subtask $n$ of device $m$ from edge server to remote cloud\\
            $E_{m,n}^{t}$ &Energy cost during transmission between mobile device and edge server for subtask $n$ of device $m$ \\
            $E_{m,n}^{r}$ &Energy cost during transmission between edge and cloud for subtask $n$ of mobile device $m$ \\
            $t_{m,n}^{l}$ & The delay when executing subtask $n$ locally\\
            $E_{m,n}^{l}$ & Energy consumption when executing subtask $n$ of device $m$\\
            $TF_{m,n}^{l}$ & Completing time of subtask $n$ on mobile device $m$ that executed locally \\
            $EF_{m,n}^{l}$ & Energy cost during the completing time of subtask $n$ on device with local computing\\
            $Budget_{m}$ & Budget or allowed delay threshold for subtasks on device $m$  \\
            $E_{m}$ & Total energy cost for all subtasks of device $m$ \\
            $TF_{m}$ &Total time consumed for all subtasks of mobile device $m$ \\
            $U_{p}^{f}$ & Profite of the edge server \\
            $X_{m,n}^{l}$ & Offloading policy for subtask $n$ of device $m$ on local computing\\
            $X_{m,n}^{f}$ & Offloading policy for subtask $n$ of  device $m$ on edge computing \\
            $X_{m,n}^{c}$ & Offloading policy for subtask $n$ of device $m$ on remote execution \\
            \hline
		\end{tabular}
	\end{center}
\end{table}

\subsection{Communication Model}
\paragraph{Transmission between mobile devices and edge.}
Let $X_{m,n}^m \in \{0,1\}$, $X_{m,n}^f \in \{0,1\}$ and $X_{m,n}^c \in \{0,1\}$ each represents the computation offloading policy made by the $m$-th mobile device. Particularly, $X_{m,n}^m = 1$ denotes that the subtask $n$ on mobile device $m$ is executed locally while $X_{m,n}^f = 1$ denotes that the subtask $n$ of mobile device $m$ is executed on the edge server. Similarly, $X_{m,n}^c = 1$ denotes that the subtask $n$ on mobile device $m$ is executed on the remote cloud. We can compute the uplink data rate for wireless transmission between mobile device and edge server as \cite{chen2018brains}:
\begin{equation}
	R_{m,n} = Wlog_2\left( 1 +  \frac{P_{m, n}^{m}G_{m, n}} { \sigma_{m}^{2}+\sum_{i\neq  m,j \neq m} P_{i,j}^{m}G_{i,j}} \right),
\end{equation}
where $P_{m, n}^{m}$ is the transmission power of mobile device $m$ to upload the subtask $n$ to the edge server via AP, $G_{m, n}$ is the channel gain between the $m$th mobile device and the corresponding AP when transmitting subtask $n$. $G_{m, n} = (dis_{m, p}^{-\eta})\left | h_{m,p} \right |^2$ where $ dis_{m, p}$ denotes the Euclidean distance between mobile device and edge server, $h_{m,p}$ is the corresponding Rayleigh fading channel coefficient that obeys the distribution of $N(0,1)$ \cite{Zhang2016Primary}. The surrounding noise power at the receiver, i.e. the AP, is $\sigma_{m}^2$ \cite{Zhang2016Primary}.

It should be noted that, for the benefit of presentation, the downlink transmisssion rate is represented by the corresponding uplink rate. In the following expressions, we also utilize the expression of uplink transmission delay to repsent the downlink transmission delay. That's because the downlink transmision rate is usually a few times larger than the uplink transmission rate due to the channel allocation result of network operator. With this change, we can reduce the complexity of delay and energy cost expressions, which will be described in detail in following paragraphs. 

The transmission delay of subtask $n$ between mobile device $m$ and the corresponding edge server thus can be \cite{Dinh2017Offloading} 
\begin{equation}
	t_{m,n}^{t} = \frac{D_{m,n}}{R_{m,n}},
\end{equation}
where $t_{m,n}^{t}$ represents the time spent on sending the subtask $n$ on mobile device $m$ to the edge server, while $D_{m,n}$ is the data size of the subtask $n$ of device $m$. Based on the above equations, we can obtain the energy consumption when transmitting subtask $n$ of mobile device $m$ to the edge server as
\begin{equation}
	E_{m,n}^{t} = P_{m,n}^{m} t_{m,n}^{t},
\end{equation}
where $P_{m,n}^{tx}$ is the power of mobile device $m$ when sending subtask $n$.
 
\paragraph{Transmission between edge and cloud.}
Due to fact that the edge server links the remote cloud via wired connection, the delay of data transmission from edge server to the cloud thus is
\begin{equation}
	t_{m,n}^{r} = \frac{D_{m,n}}{\omega},
\end{equation} 
where $t_{m,n}^{r}$ denotes the transmission delay for subtask $n$ of mobile device $m$ from edge server to the cloud. $\omega$ denotes the upstream bandwidth. Given the transmission delaybetween edge and remote cloud $t_{m,n}^{r}$ and the transmission power $P_{0}$, $E_{m,n}^{r}$ can be expressed as
\begin{equation}
	E_{m,n}^{r} = P_{0}t_{m,n}^{r},
\end{equation} 
where $E_{m,n}^{r}$ is the energy consumed when sending the subtask $n$ of mobile device $m$ from edge to the cloud.

\subsection{Computation Model}
\paragraph{Computation on local device.} Let $f_{m}^{l}$ be the CPU clock speed of mobile device $m$ and $W_{m, n}$ be the workload of subtask $n$ of mobile device $m$, if the subtask $n$ on mobile device $m$ is executed locally, then the subtask's execution time is
\begin{equation}
	t_{m,n}^{l} = \frac{W_{m,n}}{f_{m}^{l}}.
\end{equation}
Given the computation time $t_{m,n}^{l}$, the energy consumed for subtask $n$ of mobile device $m$ for local computing is
\begin{equation}
\label{equationEmnl}
	E_{m,n}^{l} = kW_{m,n}{f_{m}^{l}}^{2}.
\end{equation}
By default, $k$ is set as ${10}^{-11}$ following \cite{Guo2016Energy}. 

\paragraph{Computation on edge.} Let $f^{f}$ be the CPU frequency of edge server, if the subtask $n$ of mobile device $m$ is executed on the edge server, the computation time of the edge server can be
\begin{equation}
	t_{m,n}^{f} = \frac{W_{m,n}}{f^f},
\end{equation}
and the energy cost of edge server can be expressed as:
\begin{equation}
\label{equationEmnf}
E_{m,n}^{f} = \left( \alpha_f \left( f^f\right)^{\sigma}  + \beta_f \right) t_{m,n}^{f}.
\end{equation}
According to \cite{Rao2012Distributed}, $\alpha_f$ and $\beta_f$ are the positive constants which can be obtained by offline power fitting and $\sigma$ ranges from 2.5 to 3. If subtask $n$ of mobile device $m$ is executed on the cloud, the computation delay and energy cost of remote cloud are as follows:
\begin{equation}
    t_{m,n}^{c} = \frac{W_{m,n}}{f^c},
\end{equation}
and 
\begin{equation}
       \label{equationEmnc}
		E_{m,n}^{c} = \left( \alpha_c \left( f^c\right)^{\sigma}  + \beta_c \right) t_{m,n}^{c}.
\end{equation} 

\subsection{Dependency Constraints}
\begin{definition}
	 \textbf{Subtask's completing time:} subtask $n$ of mobile device $m$ can only start when all its predecessor subtasks has been completed. The completion time for the $n$th subtask of mobile device $m$ is consisted of two parts: the time spent to obtain the results of all its predecessor tasks and the time spent for its own computation.
\end{definition}

\begin{definition}
	\textbf{Energy cost to accomplish one subtask:} it is also consisted of two parts: the energy spent getting the result of predecessor tasks and the energy spent for its own execution.
\end{definition}

Base on the above definitions, if subtask $n$ of mobile device $m$ is assigned to be executed locally, its completion time can be expressed as:
\begin{equation}
	TF_{m,n}^{l} = max_{k \in pre\left(n\right)} \left \{  X_{m,k}^{f}t_{m,n}^{t} + X_{m,k}^{c}\left(  t_{m,n}^{t} + t_{m,n}^{r}\right)\right\}+t_{m,n}^{l}
	\label{equation_tfmn},
\end{equation}
and the energy cost for local completion is
\begin{equation}
	EF_{m,n}^{l} = \sum_{k \in pre\left(n\right) } \left[ X_{m,k}^{f} E_{m,n}^{t} + X_{m,k}^{c}\left( E_{m,n}^{t}+E_{m,n}^{r}\right)\right] + E_{m,n}^{l}
	\label{equation_efmn}.
\end{equation}
In (\ref{equation_tfmn}) and (\ref{equation_efmn}), $X_{m,k}^{f} X_{m,k}^{c}=0$. The notation $pre(n)$ in (\ref{equation_tfmn}) means all the predecessor subtasks of the $n$th subtask. In (\ref{equation_tfmn}), the term $X_{m,k}^{f}t_{m,n}^{t}$ is the delay to obtain the predecessor subtask's result of the $n$th subtask, if the predecessor subtask of $n$ is executed on the edge server. Similarly, the term $X_{m,k}^{c}\left(t_{m,n}^{t} + t_{m,n}^{r}\right)$ is the delay to obtain the result if the predecessor subtask of $n$ is accomplished on the cloud server. 
 
If subtask $n$ of mobile device $m$ is assigned to be executed on the edge server, the completion time of subtask $n$ can be defined as:
\begin{equation}
\label{tff}
	TF_{m,n}^{f} = max_{k \in pre\left(n\right)} \left\{ X_{m,k}^{m}t_{m,n}^{t} + X_{m,k}^{c}t_{m,n}^{r}\right\} + t_{m,n}^{f},
\end{equation}
where $X_{m,k}^{m}$ is predecessor subtask's assignment strategy on mobile device. $X_{m,k}^{m}=1$ means the $k$th subtask is computed on the local mobile device, while $X_{m,k}^{m}=0$, otherwise. The term $X_{m,k}^{m}t_{m,n}^{t}$ is the delay to transmit the result of predecessor task from mobile device to the edge server while $X_{m,k}^{c}t_{m,n}^{r}$ is the delay to send the prior result from the remote cloud to the edge server. 

Let $EF_{m,n}^{f}$ be the energy cost for subtask $n$ of device $m$ executed on the edge server, similarly as (\ref{equation_efmn}), it can be defined as
\begin{equation}
\label{eff}
EF_{m,n}^{f} = \sum_{k \in pre\left(n\right)} \left[X_{m,k}^{m}E_{m,n}^{t} + X_{m,k}^{c}E_{m,n}^{r}\right] + E_{m,n}^{f}.
\end{equation}

Similarly as (\ref{equation_tfmn}) and (\ref{tff}),  if subtask $n$ of mobile device $m$ is assigned to be executed in the remote cloud, its completion time can be expressed as
\begin{equation}
	TF_{m,n}^{c} = max_{k \in pre\left(n\right)} \left\{ X_{m,k}^{m} \left(t_{m,n}^{t} + t_{m,n}^{r}\right) + X_{m,k}^{f}t_{m,n}^{r}\right\} + t_{m,n}^{c},
\end{equation}
and the corresponding energy cost to complete the subtask on the remote cloud, $EF_{m,n}^{c}$ is 
\begin{equation}
\label{efc}
EF_{m,n}^{c} = \sum_{k \in pre\left(n\right)} \left[ X_{m,k}^{m}\left(E_{m,n}^{t} + E_{m,n}^{r}\right) + X_{m,k}^{f}E_{m,n}^{f}\right] + E_{m,n}^{c}.
\end{equation}

\subsection{Utility constraints}
Next, we drive the utility constraints of edge server and the time budget for the completion time. The utility of edge server is
\begin{equation}
	U_{p}^{f} = \sum_{m=1}^{M}\sum_{n=0}^{N}\left(P^{f}X_{m,n}^f - E_{m,n}^{r}X_{m,n}^{c}\right),
\end{equation}
where $U_{p}^{f}$ is the utility of the edge server, $P^{f}$ is  service price of edge server.

\section{Problem Formulation}
In this section, we will present the problem formulation with constraint of time budget and utility constraint $U_{p}^{f}$. Firstly, the completion time of all tasks on mobile device $m$ can be defined as
\begin{equation}
	TF_{m} = \sum_{n=0}^{N} \left[X_{m,n}^{m}TF_{m,n}^{l} + X_{m,n}^{f}TF_{m,n}^{f} + X_{m,n}^{c}TF_{m,n}^{c}\right],
\end{equation}
where $TF_{m,n}^{l}$ is the task completion time of subtask $n$ if it is executed locally, $TF_{m,n}^{f}$ is the task completion time of subtask $n$ if it is executed on the edge server and $TF_{m,n}^{c}$ is the task completion time of subtask $n$ if it is executed on the remote cloud. 

The total energy consumption of one application, which is denoted as $E_m$ is 
\begin{equation}
	E_{m} = \sum_{n=0}^{N} \left[EF_{m,n}^{l}X_{m,n}^{l} + EF_{m,n}^{f}X_{m,n}^{f} + EF_{m,n}^{c}X_{m,n}^{c}\right],
\end{equation}
where $EF_{m,n}^{l}$ is the energy consumption of subtask $n$ if it is executed on the mobile device, $EF_{m,n}^{f}$ is the energy cost of subtask $n$ if it is executed on edge server and $EF_{m,n}^{c}$ is the energy cost of subtask $n$ if it is executed on the remote cloud.

In this work, the goal is to minimize the total energy consumption of tasks while meeting the completion time constraint. Meawhile, the utility of the edge server $U_{p}^{f}$ is guaranteed. The energy consumption minimization problem thus can be defined as:
\begin{equation*}
\centering
\begin{split}
 \textbf{OPT-1} \quad obj:\min E_{m}
\end{split}
\end{equation*}
\begin{equation*}
\centering
\begin{split}
& \quad \quad \quad \quad   \quad \quad \quad  \quad \quad \quad  \quad \quad \quad C1:  U_{p}^{f} > 0, \\
& \quad \quad \quad \quad  \quad \quad \quad   \quad \quad \quad    \quad \quad C2:  TF_{m} < Budget_{m},  \\
&\quad \quad C3 : X_{m,n}^{m} \in \{0,1\}, X_{m,n}^{f} \in \{0,1\},X_{m,n}^{c} \in \{0,1\}, n \in [0,N], m \in [1,M],  \\
& \quad \quad \quad \quad C4:   X_{m,n}^{m}+X_{m,n}^{f}+X_{m,n}^{c}= 1,  n \in [0,N], m \in [1,M].  
\end{split}
\end{equation*}
Where constraint $C1$ is the utility constraint which guarantees the positive utility of the edge server. $C2$ is the task completion time budget, i.e., the delay constraint. $C3$ lists binary constraints and $C4$ is the unique solution constraint, which means that one subtask can only be executed at one place.

\begin{theorem}
	\label{theorm1}
	The sum task completion energy minimization problem for computation offloading in this study is NP-hard.
	\begin{proof}
We transform the oritnal problem depectied in $OPT-1$ and consider a special case that the mobile device, edge server and remote cloud server are with the same configurations, which result in the same energy costs and executing time when executing tasks. Regarding each subtask as a goods with value and weight, then the value corresponds to the execution time while the weight corresponds to the energy cost. Then we ignore the task dependency constraint between subtasks as well as the constraint $C1$. $C2$ can then be viewed as the knapsack's value constraint. Therefore, the relaxed problem of $OPT-1$ has changed into a knapsack problem \cite{Kellerer2004Knapsack} which is NP-hard. Therefore, the oritinal problem $OPT-1$ is also NP-hard, which concludes this proof.
\end{proof}
\end{theorem}

\section{Algorithms}
\subsection{Gain method}
\par Based on the above models and analysis, first of all , we design a greedy method named Gain to minimize the energy consumption of mobile device $m$ when finish executing tasks. To acquire the minimum energy cost of all subtasks in an application on mobile device $m$, the minimum energy cost of subtask $n$ is selected from $EF_{m,n}^{l}, EF_{m,n}^{f}, EF_{m,n}^{c}$. This subtask-procedure is shown between Lines 1 to 11 of Algorithm \ref{alg:Gain}.

Then, we iteratively adjust the initial offloading policy to fit for the constraint of $U_{p}^{f}$ and the completion time budget $Budget_{m}$. If the offloading policy does not satisfy the constraint of $U_{p}^{f}$, which means that the number of subtask executed on remote cloud is too much to make the edge server get profits when serving mobile users. To fit the constraint of $U_{p}^{f}$, we must offload some subtasks from the remote cloud to mobile device or to the edge servers.

Then the algorithm chooses subtask considering which subtask will be offloaded. To obtain the minimum energy cost, we take the changing energy cost as the criteria to set the priority. The smaller the changing energy cost is, the higher the priority will be. 

To fit for the constraint of completion time budget, we compute the changing completion time  and the changing energy cost  in  each offloading choice. We choose the corresponding offloading stategy in the choice, which decreases the changing completion time and guarantees the minimum changing of energy cost. Due to the constraint of utility $U_{p}^{f}$, the choosing of offloading site for subtasks should be very careful. If subtask $n$ is assigned to be executed on mobile device, the offloading choice must be from mobile device to the edge server. If subtask $n$ is assigned to be executed on edge server, the offloading choice must be from edge server to mobile device. If subtask $n$ is assigned to be executed on remote cloud, the offloading choice can either be from the remote cloud to edge serve or from the remote cloud to mobile device. The detail of the Gain algorithm is depicted in Algorithm \ref{alg:Gain}.

\begin{algorithm}[htbp]
	\caption{Gain method for mobile device $m$}
	\label{alg:Gain}
	\begin{algorithmic}[1]
		\REQUIRE ~~\\
		$tasks$: a sequence of $N$ subtask-tasks mobile device \\ $m$, the execute order of subtasks;  \\
		$W$: the workload size of subtasks;  \\
		$D$: the data size of subtasks;  \\
		$Budget_{m}$: the completion time budget for subtasks; \\
		$pre$: 2-D array for each subtask's predecessor's task; \\
		\ENSURE ~~\\
		$X^{m}$: the policy of subtask executed on mobile device locally;
		$X^{f}$: the policy of subtask executed on edge server; 
		$X^{c}$: the policy of subtask executed on remote cloud;
		\FOR{$n\quad in \quad tasks$}
		\STATE computer $EF_{m,n}^{l}$, $EF_{m,n}^{f}$, $EF_{m,n}^{c}$ by Equation (\ref{equation_efmn}), (\ref{eff}), (\ref{efc}) \\
		\IF { $min\left(EF_{m,n}^{l}, EF_{m,n}^{f}, EF_{m,n}^{c}\right)  = EF_{m,n}^{l}$ }
		\STATE $X_{m,n}^{l}$ $\leftarrow$ 1, $X_{m,n}^{f}$ $\leftarrow$ 0, $X_{m,n}^{c} $ $\leftarrow$ 0 
		\ENDIF
		\IF { $min\left(EF_{m,n}^{l}, EF_{m,n}^{f}, EF_{m,n}^{c}\right)  = EF_{m,n}^{f}$ }
		\STATE $X_{m,n}^{l}$ $\leftarrow$ 0, $X_{m,n}^{f}$ $\leftarrow$ 1, $X_{m,n}^{c} $ $\leftarrow$ 0 
		\ENDIF
		\IF { $min\left(EF_{m,n}^{l}, EF_{m,n}^{f}, EF_{m,n}^{c}\right)  = EF_{m,n}^{c}$ }
		\STATE $X_{m,n}^{l}$ $\leftarrow$ 0, $X_{m,n}^{f}$ $\leftarrow$ 0, $X_{m,n}^{c} $ $\leftarrow$ 1
		\ENDIF
		\ENDFOR
		\STATE compute $U_{p}^{f}$ and $TF_{m}$
		\WHILE {$U_{p}^{f} \leq 0 \parallel  TF_{m} \geq Budget_{m}$ }
		\IF {$U_{p}^{f} \leq 0$}
		\STATE choose the subtask that bings about minimum changing energy consumption when offloading the subtask from the remote cloud to the edge server, or from the remote cloud to mobile device. 
		\ENDIF
		\IF {$TF_{m} \geq Budget_{m}$}
		\FOR{$n=0 \to N$}
		\IF {$X_{m,n}^{m} = 1$}
		\STATE compute the changing energy cost when offloading the subtask from mobile device to the edge server.
		\ENDIF
		\IF {$X_{m,n}^{f} = 1$}
		\STATE compute the changing energy cost when offloading the subtask from the edge server to mobile device
		\ENDIF
		\IF {$X_{m,n}^{c} = 1$}
		\STATE compute the changing energy cost when offloading the subtask from remote cloud to mobile device or from remote cloud to edge server.
		\ENDIF
		\STATE choose the offloading policy with the minimum changing energy cost and decrease changing completing time
		\ENDFOR
		\ENDIF
		\ENDWHILE
	\end{algorithmic}
\end{algorithm}
\begin{theorem}
	\label{theorm2}
	The time complexity of the Gain algorithm is $O(N)$.
	\begin{proof}
In algorithm \ref{alg:Gain}, the time complexity of subprocess from line 1 to 12 is $O(N)$ and the time complexity of subprocess from line 14 to 31 is $O(N)$ for the reason that the adjust time of time won't be more than $N$.
\end{proof}
\end{theorem}
\par Let $L_{opt}\left(X^{*}\right)$ be the the optimal energy cost, where corresponding optimal strategy is denoted as $X^{*}$. Let $L(\hat{X})$ be the energy cost of by executing Algorithm \ref{alg:Gain} and $\hat{X}$ be the strategy of Algorithm \ref{alg:Gain}. Then we have $L_{opt}\left(X^{*}\right) < L(\hat{X})$.
\begin{theorem}
	\label{theorm3}
	$ L(\hat{X}) < L_{opt}\left(X^{*}\right)(1 + \varepsilon)$
	\begin{proof}
	 As shown in  (\ref{equationproof1}), the energy cost of the strategy $L_{\hat{X}}$ can be defined as the sum of energy cost of each subtask $EF_{n}$. 
		\begin{equation}
		    \label{equationproof1}
		    L(X) = \sum_{n=0}^{N}EF_{n}
		\end{equation}
		It stands to reason that the optimal strategy always include the offloading strategy with  minimum energy cost, so the optimal energy cost can be defined as:
		\begin{equation}
		\label{proofequation2}
		L_{opt}(X^{*}) = \sum_{n=0}^{N}EF_{n}^{min}
		\end{equation}
		where $EF_{n}^{min}$ denotes the minimum  energy cost for the $n$-th subtask. It also stands to reason that the Algorithm \ref{alg:Gain} not always includes the offloading strategy with maximum energy cost, so the (\ref{proofequation3}) is true.
		\begin{equation}
			\label{proofequation3}
			L(\hat{X}) < \sum_{n=0}^{N}EF_{n}^{max}
		\end{equation}
		$EF_{n}^{max}$ denotes the maximum energy cost for the $n$-th subtask. If\quad$\sum_{n=0}^{N}EF_{n}^{max} < (1 + \varepsilon)\sum_{n=0}^{N}EF_{n}^{min}$ is true that $ L(\hat{X}) < L_{opt}\left(X^{*}\right)(1 + \varepsilon)$ is true as well. For the energy cost of each subtask , it includes the part of the energy cost of requesting the result from predecessor subtasks $et_{m,n}$ and the part of the energy cost of executing the subtask $ee_{m,n}$. so  \quad$\sum_{n=0}^{N}EF_{n}^{max} < (1 + \varepsilon)\sum_{n=0}^{N}EF_{n}^{min}$ can be rewritten as
		\begin{equation}
			\label{proofequation4}
			\sum_{n=0}^{N}(ee_{m,n}^{max} + et_{m,n}^{max}) < (1+\varepsilon)(ee_{m,n}^{min} + et_{m,n}^{min})
		\end{equation}
	    Base on Equations (\ref{equationEmnl}) (\ref{equationEmnf}) (\ref{equationEmnc}), we obtain $ee_{m,n}^{max}$ from the maximum value of $E_{m,n}^{l}$, $E_{m,n}^{f}$ and $E_{m,n}^{c}$. Similarly, we also obtain $ee_{m,n}^{min}$. Based on the  (\ref{equation_efmn}) (\ref{eff}) (\ref{efc}),  we obtain $et_{m,n}^{min}$ from the minimum value of $EF_{m,n}^{l}$, $EF_{m,n}^{f}$ and $EF_{m,n}^{c}$. Similarly,  we also obtain $et_{m,n}^{max}$. Base on the above, we can know that $0 < \varepsilon < 1$ and the  (\ref{proofequation4}) is true. So
	    $\sum_{n=0}^{N}EF_{n}^{max} < (1 + \varepsilon)\sum_{n=0}^{N}EF_{n}^{min}$  is true and  $ L(\hat{X}) < L_{opt}\left(X^{*}\right)(1 + \varepsilon)$ is true.

	\end{proof}
\end{theorem}

\subsection{Simulated annealing}
The simulated annealing (SA) \cite{Welsh1988SIMULATED} is a local search algorithm. In the basic SA, there is always a randomly selected solution in the begin. But in our algorithm, we obtain the initial solution from the Gain algorithm. Next, we initialize the temperature of SA. Then when the temperature is greater than $0.1$, we take the subprocess of adjusting the offloading policy in random and determine if the SA algorithm accepts the offloading policy. The detail of the SA algorithm is shown in Algorithm \ref{alg:SA}.

\begin{algorithm}[!htbp]
	\caption{Simulated annealing for mobile device $m$}
	\label{alg:SA}
	\begin{algorithmic}[1]
		\REQUIRE ~~\\
		$tasks$: a sequence of $N$ subtask-tasks mobile device \\ $m$, the execute order of subtasks;  \\
		$W$: the workload size of subtasks;  \\
		$D$: the data size of subtasks;  \\
		$Budget_{m}$: the completing time budget for subtasks; \\ 
		$pre$: 2-D array for each subtask's predecessor's task; \\
		$T_{0}$: initial temperature; \\
		$cool$: the speed of cooling; \\
		\ENSURE ~~\\
			$X^{m}$: the policy of subtask  executed on mobile device locally;
		$X^{f}$: the policy of subtask executed on edge server; 
		$X^{c}$: the policy of subtask executed on remote cloud;
		\STATE get the initial offload policy $X_{m}^{m}, X_{m}^{f}, X_{m}^{c}$ from gain method
		\STATE $Tem = T_{0}$
		\WHILE {$Tem > 0.1$}
		\STATE compute $EF_{m}^{1}$ based on the offload policy  $X_{m}^{m}, X_{m}^{f}, X_{m}^{c}$
		\STATE randomly choose a subtask $n$
		\STATE $TempX_{m}^{m} \leftarrow X_{m}^{m}, TempX_{m}^{f} \leftarrow X_{m}^{f}, TmepX_{m}^{c} \leftarrow X_{m}^{c}$
		\STATE $change \leftarrow random\left[1,3\right]$
		\IF {$change = 1$}
		\STATE $X_{m,n}^{l}$ $\leftarrow$ 1, $X_{m,n}^{f}$ $\leftarrow$ 0, $X_{m,n}^{c} $ $\leftarrow$ 0 
		\ENDIF
		\IF {$ change = 2$}
		\STATE  $X_{m,n}^{l}$ $\leftarrow$ 0, $X_{m,n}^{f}$ $\leftarrow$ 1, $X_{m,n}^{c} $ $\leftarrow$ 0 
		\ENDIF
		\IF {$ change = 3$}
		\STATE $X_{m,n}^{l}$ $\leftarrow$ 0, $X_{m,n}^{f}$ $\leftarrow$ 0, $X_{m,n}^{c} $ $\leftarrow$ 1
		\ENDIF
		\STATE compute $U_{p}^{f}$, $EF_{m}$and $TF_{m}$
		\IF {$U_{p}^{f} > 0 \& TF_{m} < Budget_{m}$}
		\STATE accept with probility $exp\left(\frac{-(EF_{m} - EF_{m}^{1})}{Tem}\right)$
		\STATE $X_{m}^{m} \leftarrow TempX_{m}^{m}, X_{m}^{f} \leftarrow TempX_{m}^{f}, X_{m}^{c} \leftarrow TempX_{m}^{c}$
		\ENDIF
		\ENDWHILE 
	\end{algorithmic}
	
\end{algorithm}

\section{Performance  Evaluation}
To study the performance of proposed algorithms, 
We implement the algorithms on a high performance work station with an Intel I7 processor at frequency $3.9GHz$ and has a 8G RAM. We use Python $3.6$ \cite{Aksimentiev2007Python} to simulate the offloading of subtasks and evaluate the algorithms in terms of running time, application completion time and energy cost with $100$ repeated trials. 

\begin{figure}[!h]
	\centering
	\includegraphics[width=2.5in]{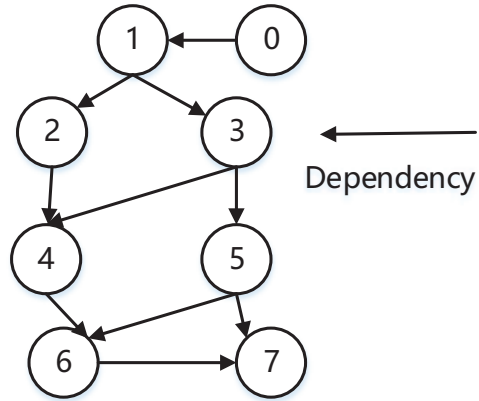}
	\caption{The task graph.} \label{task graph}
\end{figure}

In order to simulate real-world tasks, we use a typical task graph \cite{Shi2016Algorithmic} as shown in Fig. \ref{task graph}. In Fig. \ref{task graph}, dependency constraints exists between subtasks, which determine the execution order. Based on the task graph, one possible execution sequence for subtasks is $[0, 1, 2, 3, 4, 5, 6, 7]$. 
\subsection{Simulation Setup}

\par  We set the 8 subtasks with evenly distributed workload and evenly distributed data size. The signal noise between
the edge server and mobile device is set as $\sigma ^{2}=1$, the wireless bandwidth of upload is set as $W = 2Mbps $ and the wireless bandwidth of download is set as $W = 10Mbps$ \cite{Ding2010Distributed}. The bandwidth between edge server and remote cloud of upload is $W = 1024Mbps$ and the bandwidth between edge server and remote cloud of download is $W = 8192Mbps$ \cite{Ding2010Distributed}. The CPU frequency of mobile device is $f^m = 5\times10^6Hz$, while the CPU frequency of edge server is $f^f = 2\times10^9Hz$ \cite{ISI:000283943700118}. The CPU frequency of remote cloud is set as $f^c = 4\times10^9Hz$ \cite{ISI:000283943700118}. System
parameters $\alpha_f=0.1, \beta_f=0.1, \alpha_c = 0.2, \beta_c = 0.2$ \cite{ISI:000283943700118}. The communication chip power of mobile device is $0.1$watt \cite{Liu2009Dual}. The communication chip power of edge server is $1$watt \cite{Liu2009Dual} and the communication chip power of remote cloud is $3$watt \cite{Liu2009Dual}.

\subsection{Algorithm Performance}
\begin{figure}[!h]
\centering
\includegraphics[width=3in]{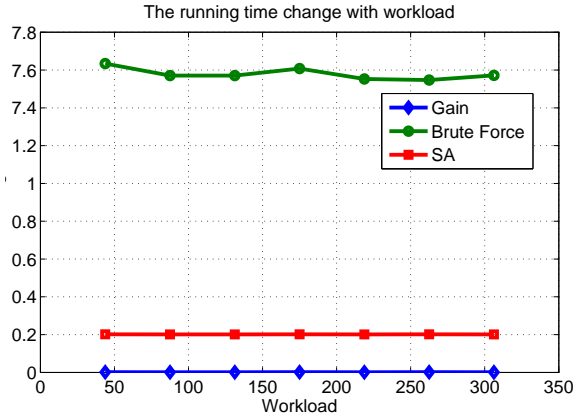}
\caption{The comparison of three algorithm's executing time on different average workload size.}
\label{f1}
\end{figure}
Fig. \ref{f1} shows the comparison of Gain, Brute Force and SA in terms of
running time with different workload sizes. From Fig. \ref{f1}, we observe that, the running time of Brute Force is range from 7.54s to 7.68s while the running time of SA is almost $0.2s$ and the running time  of Gain is less than 0.02s. This is because of that the Brute Force tries to exhaustively all solutions and  the solution space of the problem is $N^3$, where $N$ denotes the number of subtasks. From Fig. \ref{f1}, we can observe that, the running time of three algorithms almost no fluctuations, which indicates the robustness of algorithms.  In Brute Force, the maximum running time  is $7.66s$, while the minimum running time  is $7.547s$, the differential value of the maximum running time and the minimum running time is only $0.12s$. In Gain, the maximum running time is $0.0015$ and the minimum running time is $0.001$. In SA, the running time is almost $0.20$.
 

\begin{figure}[!h]
	\begin{minipage}[t]{0.49\textwidth}
		\centering
		\setlength{\belowcaptionskip}{-1em}
		\includegraphics[width=5.5in]{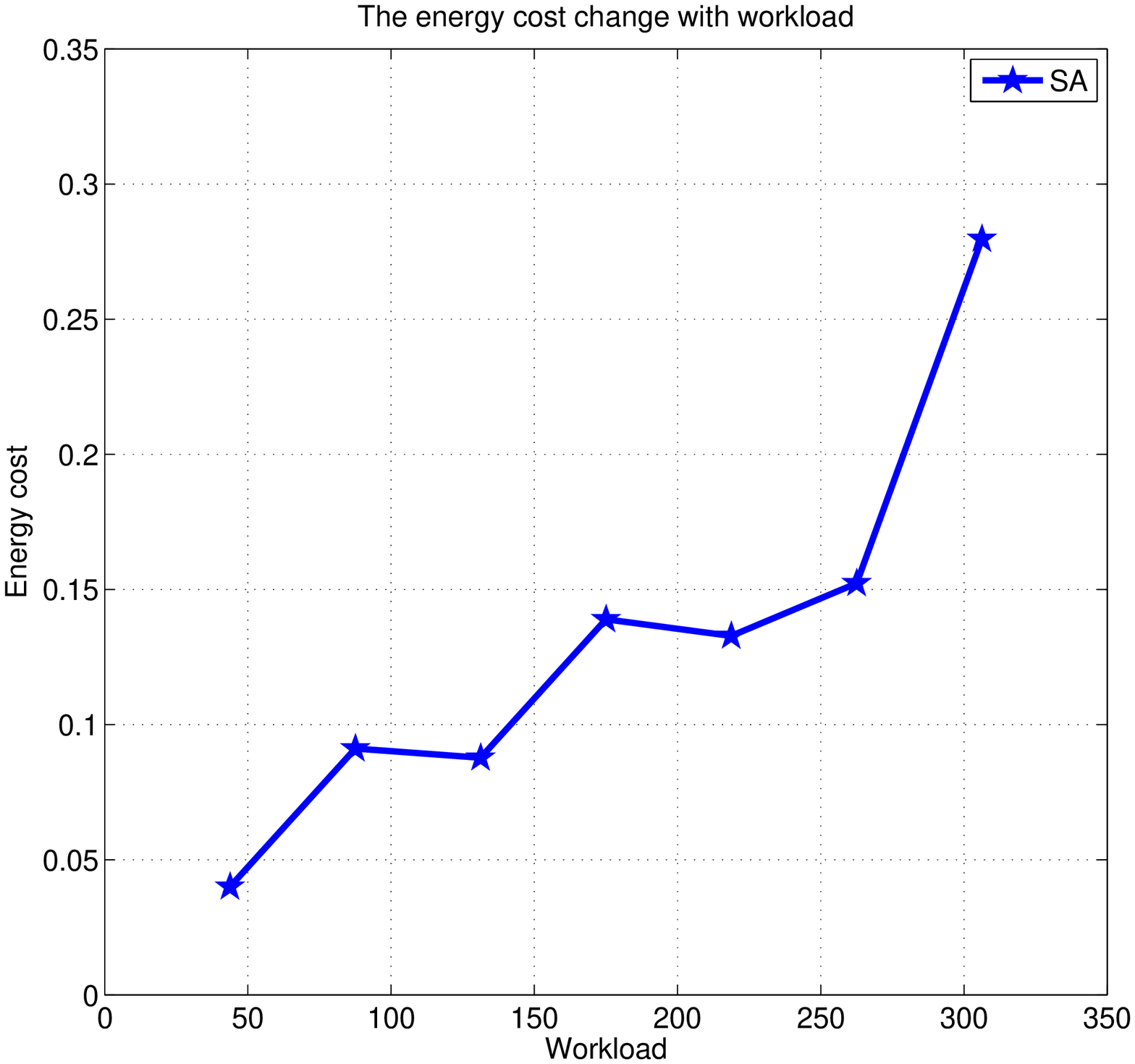}\\
		\caption{The energy cost of SA with the change of workload size.}\label{f2-1}
	\end{minipage}
	\begin{minipage}[t]{0.49\textwidth}
		\centering
		\setlength{\belowcaptionskip}{-1em}
		\includegraphics[width=5.5in]{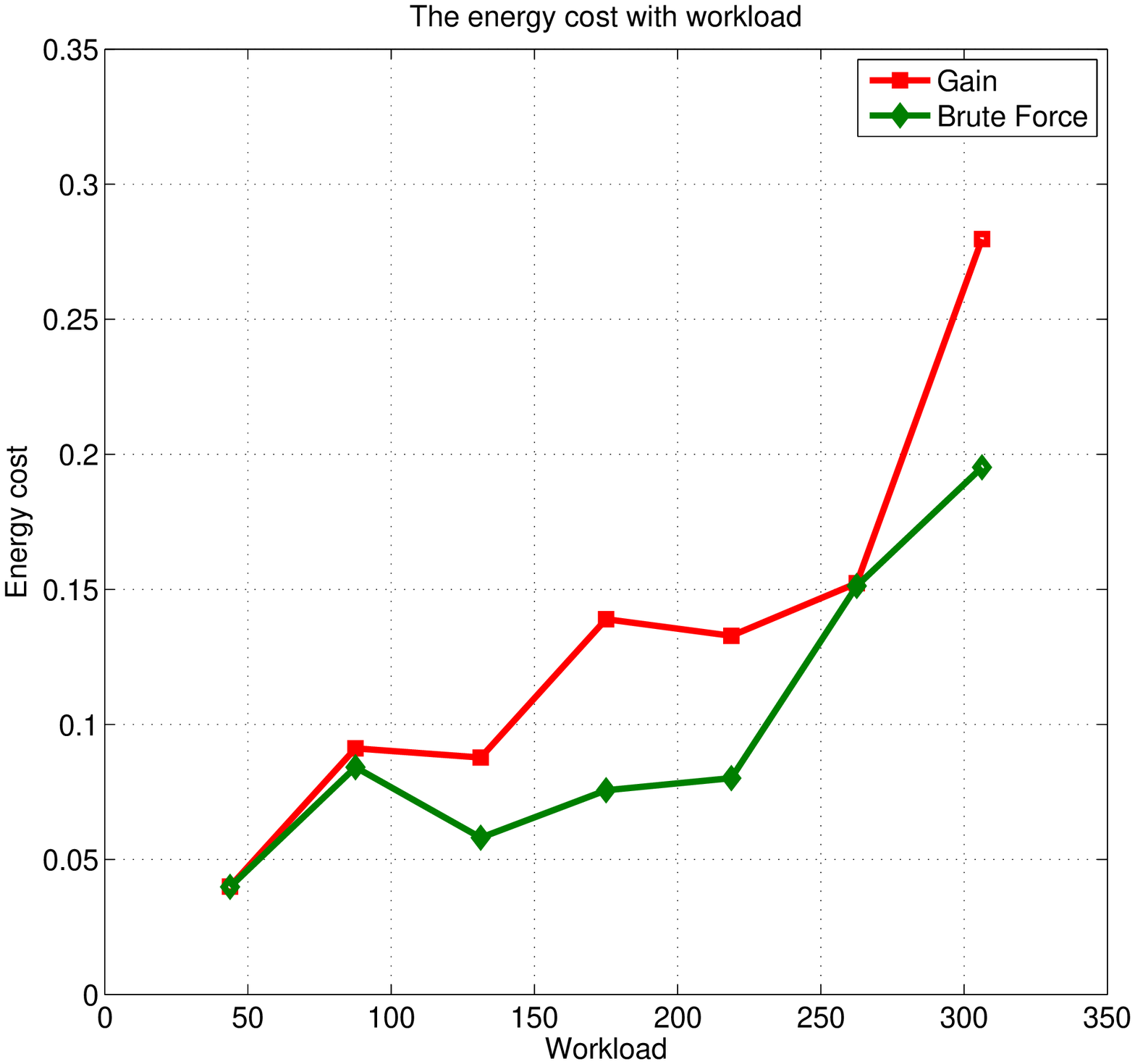}\\
		\caption{The energy cost of Gain and Brute Force with the change of workload size.}\label{f2-2}
	\end{minipage}
\end{figure}

Fig. \ref{f2-1} and Fig. \ref{f2-2} show the comparison of Gain, Brute Force and SA on  energy cost 
with different workload size. In Fig. \ref{f2-2} and Fig. \ref{f2-1}, The Brute Force always obtains the minimum energy cost compared with two other algorithms. On the other hand, because SA uses the results of Gain to initialize its initial solution and can not find more effective offloading strategy, SA always obtains the same result as Gain. From the  comparison of Brute Force and Gain, we  observe
that Gain can optimally achieve the same completion time budget performance of optional result with only $0.6\%$ extra energy cost. 
In Fig. \ref{f2-2}, when the workload size has grown from $43.75$ to $87.5$, the energy cost also increases by  $0.06$. but the energy cost falls by $0.04$ when the workload size has grown from $87.5$ to $131.25$ due to the constraint of task dependency. From Fig. \ref{f2-2}, the trends in energy consumption of Gain almost the same as the trends in energy consumption of Brute Force which indicates that The Gain always goes in the direction of the optimal solutions.

\begin{figure}[!h]
	\begin{minipage}[t]{0.49\textwidth}
		\centering
		\setlength{\belowcaptionskip}{-1em}
		\includegraphics[width=5.5in]{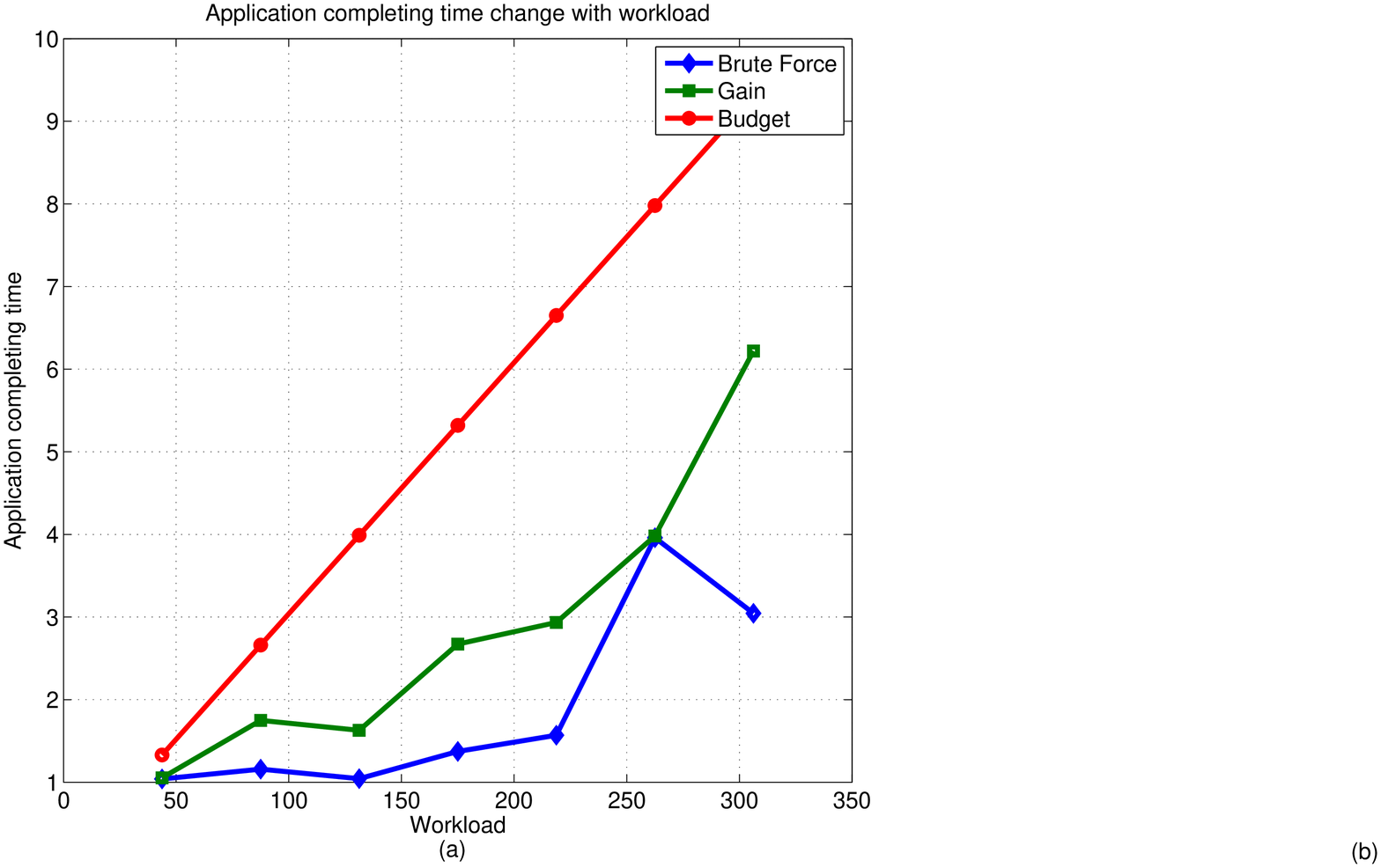}\\
		\caption{ The comparison of application completion time of Gain, Brute Force and Budget on different  workload size.}\label{f3-1}
	\end{minipage}
	\begin{minipage}[t]{0.49\textwidth}
		\centering
		\setlength{\belowcaptionskip}{-1em}
		\includegraphics[width=5.5in]{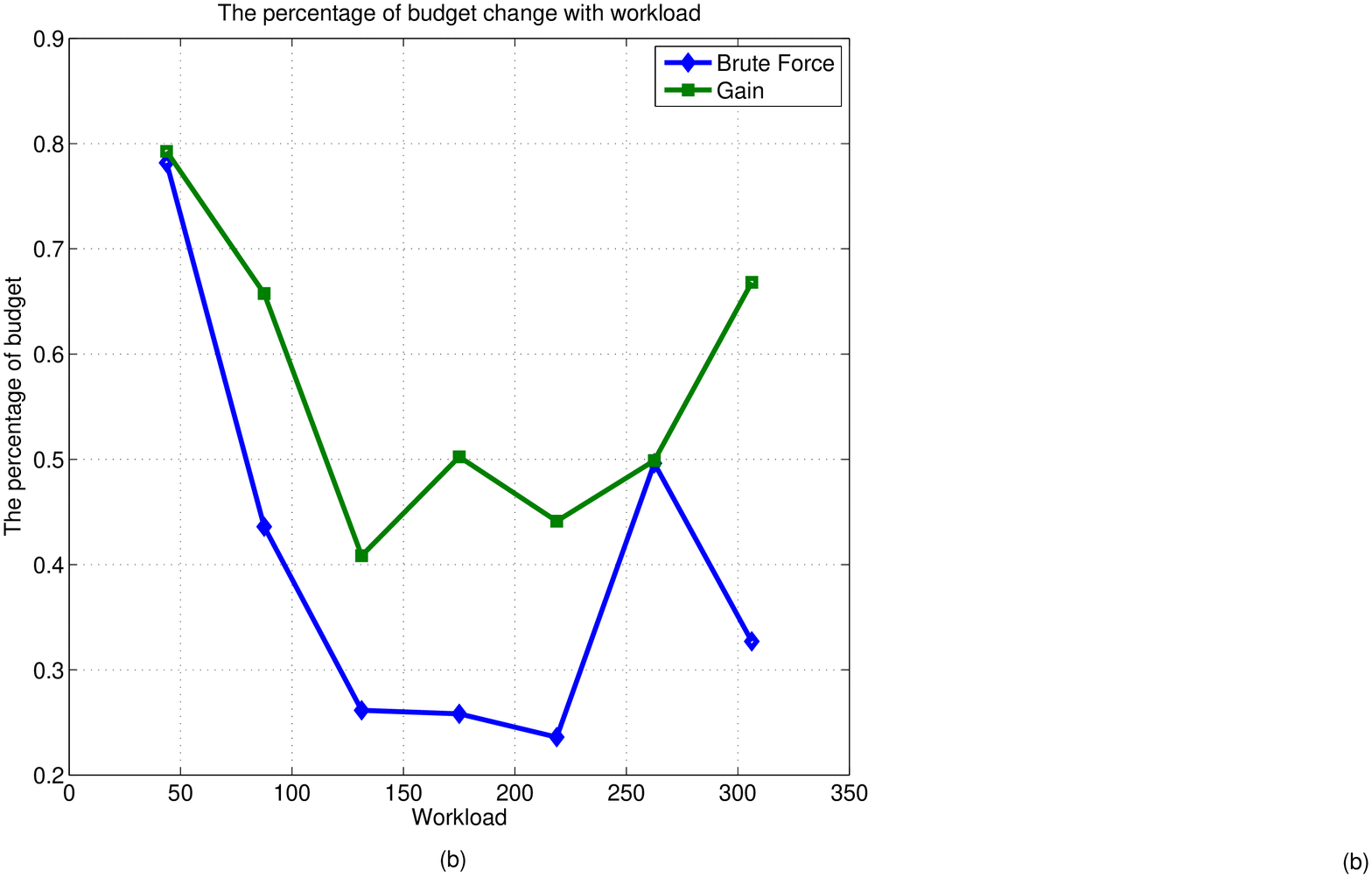}\\
		\caption{The comparison of application completion time as a percentage of Budget.}\label{f3-2}
	\end{minipage}
\end{figure}
\par Fig \ref{f3-1} shows the comparison of application completing time of Gain, Brute Force. The completion time budget $Budget$ can be represent as  (\ref{equationBduget}) and $W$ denotes the workload matrix, $N$ denotes the number of subtask of the mobile device $m$. 
\begin{equation}
	Budget = 0.5\times\sum_{n=0}^{N}W_{m,n} 
	{\label{equationBduget}}
\end{equation}
\par From Fig \ref{f3-1}, we observed that Gain and Brute Force always  obtain  a  efficient solution which satisfies the constraints. In Fig \ref{f3-2}, the completion time of Gain is range from $0.2$ to $0.8$  of the completion time budget.

\begin{figure}[!h]
	\begin{minipage}[t]{0.48\textwidth}
		\centering
		\setlength{\belowcaptionskip}{-1em}
		\includegraphics[width=5.5in]{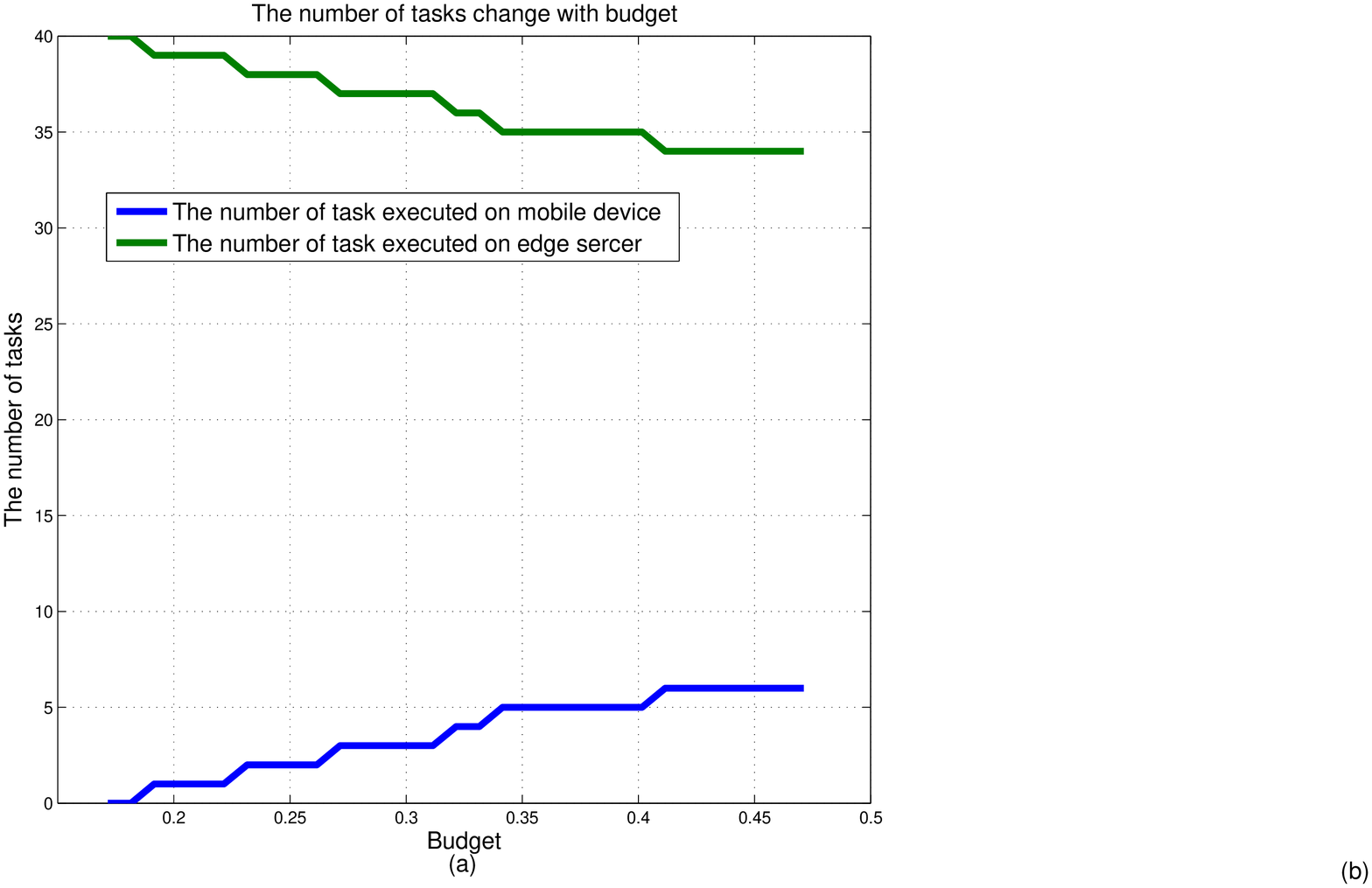}\\
		\caption{ The change of  the number of task on mobile device and edge server along with the increase of budget.}\label{f4-1}
	\end{minipage}
	\begin{minipage}[t]{0.48\textwidth}
		\centering
		\setlength{\belowcaptionskip}{-1em}
		\includegraphics[width=5.5in]{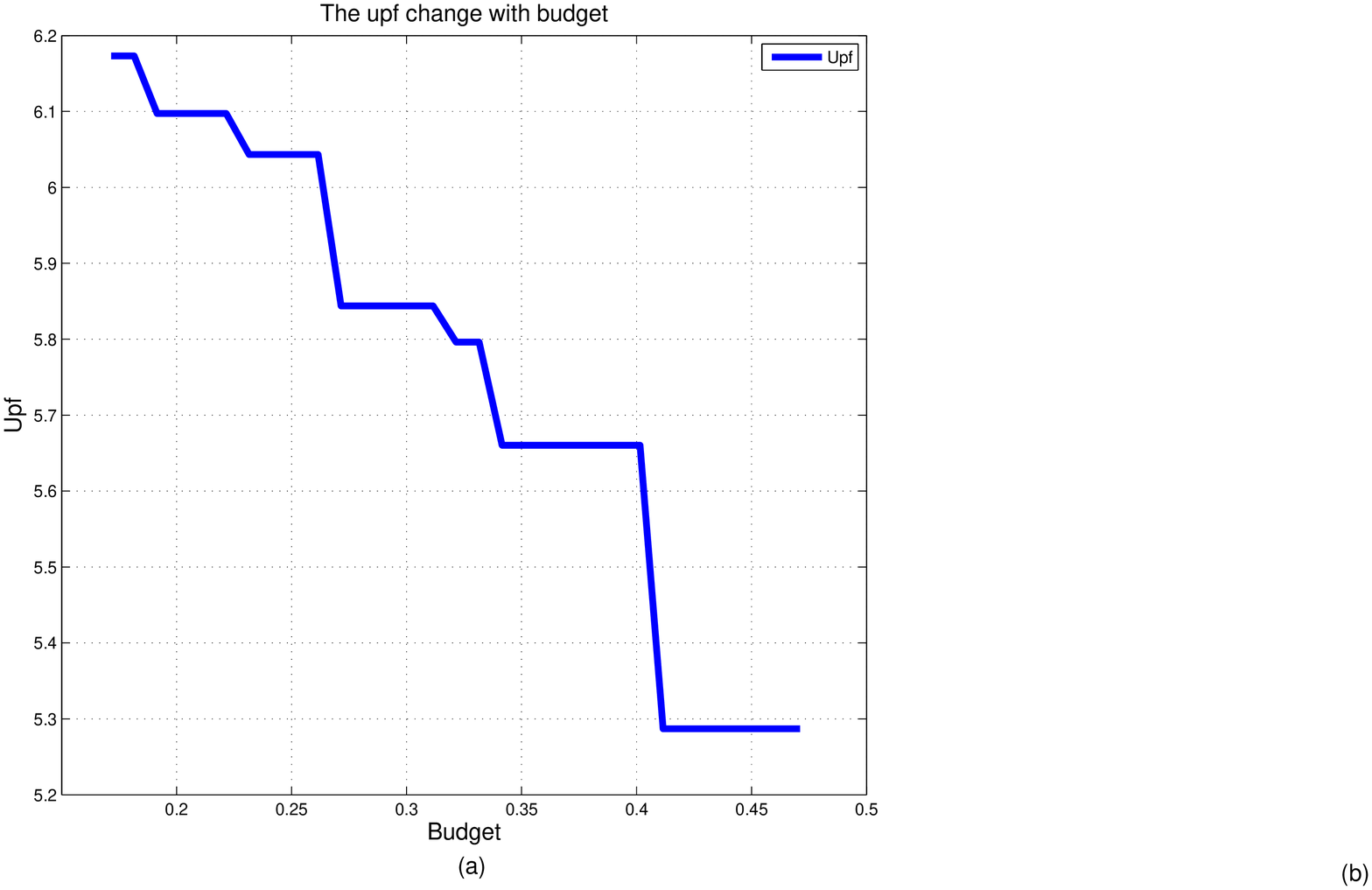}\\
		\caption{The change of $U_p^f$ with the increase of budget.}\label{f4-2}
	\end{minipage}
\end{figure}

From Fig \ref{f4-1}, along with the growth of completing time budget, the number of the task executed on edge server decrease from 40 to 34 and the number of budget executed on mobile device increases from 0 to 6. Finally, they achieve balance. it is the reason of the constraint of $U_{p}^{f}$ and the design of algorithm. In Fig \ref{f4-2}, the number of tasks assigned to be executed on edge server changes from 40 to 34 , the profile of edge server is smaller.

\section{Conclusions}
This paper has addressed novel computation offload schemes with device, fog and remote cloud collaboration. We have formulated the offload problem as a energy cost minimize problem with application completing time budget and fog profit’s constraint. The problem is NP-hard problem. Focus on that problem, we design a greedy algorithm aimed to minimize the the energy cost, which also fit the constraint of completion time,  utility and task dependency. After analyzing the results of the experiment, the following  points are obtained. Firstly, the implementation shows that in a three-tier structure such as Mobile, Edge Server and Remote cloud, Edge server plays a very important role in effectively reducing the energy consumption during task execution. Secondly, the proposed greedy algorithm can achieve the same application completing time budget performance of the Brute Force optional algorithm with only 31
cost. The simulated annealing algorithm can achieve similar performance
with the greedy algorithm.
\par In the future, we will devise online algorithms by modifying the initialization process of each algorithms and discuss the min energy cost problem with each subtask’s completing time budget.

\section{Acknowledgment}
This work was supported by the National Key R\&D Program of China under Grant No. 2018YFB1003201. It was also supported in part  by the National Natural
Science Foundation of China under Grant Nos. 61702115 and
61672171, by the Major Research Project of Educational Commission
of Guangdong Province under Grant No. 2016KZDXM052, and the
China Postdoctoral Science Foundation Fund under Grant No.
2017M622632.

\bibliographystyle{splncs04}
\bibliography{ref}

\end{document}